\documentclass[aip, superscriptaddress, reprint, twocolumn]{revtex4-1}
\usepackage{natbib}
\usepackage{graphicx}
\usepackage{wrapfig} % to wrap text around pictures
\usepackage{color}
\usepackage{hyperref}

\begin{document}

\title{
Citation inequity and gendered citation practices in contemporary physics
}

\author{Erin G. Teich}
\affiliation{Department of Bioengineering, School of Engineering \& Applied Science, University of Pennsylvania, Philadelphia, PA 19104 USA}

\author{Jason Z. Kim}
\affiliation{Department of Bioengineering, School of Engineering \& Applied Science, University of Pennsylvania, Philadelphia, PA 19104 USA}

\author{Christopher W. Lynn}
\affiliation{Initiative for the Theoretical Sciences, Graduate Center, City University of New York, New York, NY 10016, USA}
\affiliation{Joseph Henry Laboratories of Physics and Lewis–Sigler Institute for Integrative Genomics, Princeton University, Princeton, NJ 08544, USA}

\author{Samantha C. Simon}
\affiliation{Department of Physics \& Astronomy, College of Arts \& Sciences, University of Pennsylvania, Philadelphia, PA 19104 USA}

\author{Andrei A. Klishin}
\affiliation{Department of Bioengineering, School of Engineering \& Applied Science, University of Pennsylvania, Philadelphia, PA 19104 USA}

\author{Karol P. Szymula}
\affiliation{School of Medicine and Dentistry, University of Rochester, Rochester, NY 14642 USA}

\author{Pragya Srivastava}
\affiliation{Department of Bioengineering, School of Engineering \& Applied Science, University of Pennsylvania, Philadelphia, PA 19104 USA}

\author{Lee C. Bassett}
\affiliation{Department of Electrical \& Systems Engineering, School of Engineering \& Applied Science, University of Pennsylvania, Philadelphia, PA 19104 USA}

\author{Perry Zurn}
\affiliation{Department of Philosophy \& Religion, American University, Washington, D.C. 20016 USA}

\author{Jordan D. Dworkin}
\affiliation{Department of Psychiatry, Columbia University College of Physicians and Surgeons, New York, NY 10019 USA}
\affiliation{New York State Psychiatric Institute, New York, NY 10032 USA}

\author{Dani S. Bassett}
\affiliation{Department of Physics \& Astronomy, College of Arts \& Sciences, University of Pennsylvania, Philadelphia, PA 19104 USA}
\affiliation{Department of Bioengineering, School of Engineering \& Applied Science, University of Pennsylvania, Philadelphia, PA 19104 USA}
\affiliation{Department of Electrical \& Systems Engineering, School of Engineering \& Applied Science, University of Pennsylvania, Philadelphia, PA 19104 USA}
\affiliation{Department of Neurology, Perelman School of Medicine, University of Pennsylvania, Philadelphia, PA 19104 USA}
\affiliation{Department of Psychiatry, Perelman School of Medicine, University of Pennsylvania, Philadelphia, PA 19104 USA}
\affiliation{Santa Fe Institute, Santa Fe, NM 87501 USA}
\affiliation{To whom correspondence should be addressed: dsb@seas.upenn.edu}

\begin{abstract} 
The historical and contemporary under-attribution of women's contributions to scientific scholarship is well-known and well-studied, with effects that are felt today in myriad ways by women scientists.
One measure of this under-attribution is the so-called citation gap between men and women: the under-citation of papers authored by women relative to expected rates coupled with a corresponding over-citation of papers authored by men relative to expected rates.
We explore the citation gap in contemporary physics, analyzing over one million articles published over the last 25 years in 35 physics journals that span a wide range of subfields.
Using a model that predicts papers' expected citation rates according to a set of characteristics separate from author gender, we find a global bias wherein papers authored by women are significantly under-cited, and papers authored by men are significantly over-cited.
Moreover, we find that citation behavior varies along several dimensions, such that imbalances differ according to \emph{who} is citing, \emph{where} they are citing, and \emph{how} they are citing.
Specifically, citation imbalance in favor of man-authored papers is highest for papers authored by men, papers published in general physics journals, and papers likely to be less familiar to citing authors.
Our results suggest that, although deciding which papers to cite is an individual choice, the cumulative effects of these choices needlessly harm a subset of scholars.
We discuss several strategies for the mitigation of these effects, including conscious behavioral changes at the individual, journal, and community levels.
\end{abstract}

\date{\today}

\maketitle

\section{Introduction}

The under-attribution of women's contributions to academic scholarship in general---and scientific scholarship in particular---has been recognized for over 150 years \cite{Gage1870, Rossiter1993}.
A broad body of work has studied this so-called ``Matilda effect" \cite{Rossiter1993} of under-attribution using a variety of tools ranging from feminist theory \cite{Phillips2000} to statistics \cite{Caplar2017}.
The devaluing of women's contributions in academia manifests in a decremented interest in collaborating with women \cite{Knobloch-Westerwick2013}, the devaluation of women's contributions to scientific research articles \cite{chaoqun2021gendered}, a pervasive perception that women have less academic excellence and ability \cite{Knobloch-Westerwick2013, VandenBrink2012, Moss-Racusin2012, Bloodhart2020}, and a marked dearth of academic awards given to women \cite{Lincoln2012, Lunnemann2019}. 
Women also face longer publication review processes \cite{hengel2020publishing}, fewer invited paper commissions \cite{Holman2018}, and lower citation rates \cite{Sugimoto2013, Ghiasi2015, Caplar2017, Dworkin2020, Wang2021, Dion2018, Maliniak2013, Mitchell2013, Fulvio2020, Chatterjee2021}.
Moreover, the under-attribution of women's academic accomplishments represents a dangerous erasure of women from the collectively built story of the history and future of scientific progress.
Exclusion of women from scientific textbooks \cite{Phillips2000, Damschen2005, Wood2020a, Simpson2021}, for example, contributes to the construction of damaging stereotypes regarding who can be a good scientist \cite{Carli2016, McKinnon2020}.
Internalization of these stereotypes can impact scientific performance \cite{Good2010a, Schinske2016} and interest \cite{Cheryan2015, Schinske2016} among women and girls, leading to greater exclusion still.
Exposure of the under-attribution of women's contributions to science, followed by mitigation, is thus a means of interrupting a historical cycle of exclusion and hegemony within the sciences. 
Throughout, rather than assuming individual intent towards exclusion, we take a consequentialist and post-intent approach that is focused on the nature of current cultural and social practices, and the collective effects of actions, in the spirit of other work addressing structural oppression \cite{Kendi2019,perry2006post}.

Here, we focus on quantifying the pervasive under-citation of woman authors throughout the field of physics, and discuss ways in which citation inequity may be mitigated by individual authors and journals.
We use a proxy for author gender---namely, a statistical correlation between the forename under which an author publishes and gender identity, determined from two public databases---in all analyses.
We utilize this proxy both because the data it relies on is immediately available and analyzable via automated means, and also (importantly) because names have a great deal of influence over individual perceptions of authors' gender identities \cite{Pilcher2017} and thus over individual assumptions (implicit or explicit) regarding the merit of authors' scientific works. 
This influence exists regardless of those authors' self-attested, actual gender identities.
We show that the under-citation of women('s names) does not exist diffusely as a background effect baked into the fabric of publication in physics, but rather depends on \textbf{citation venue}, \textbf{citation actor}, and \textbf{form of citation}.

We analyze over one million papers published in 35 physics journals between 1995 and 2020, and examine citation practices of these papers as a function of name-based author gender category, physics subfield, and publication date.
We find that, even after controlling for paper characteristics that are separate from author gender category (including year and journal of publication), there is a global imbalance in favor of man-authored papers.
This citation inequity is especially strong for citing papers within the broad category of ``general" physics (\textbf{citation venue}).
It is driven primarily by the citing behavior of man authors, whose imbalanced citation practice remains stable over time (\textbf{citation actor}).
Finally, global citation behavior trends more toward the over-citation of man-authored papers when citations refer to work with which citing authors may be less familiar (\textbf{citation form}). 
This effect is driven primarily by man author teams, who exhibit citation preference for other man author teams when citing both familiar and less familiar work. 
In contrast, among woman authors we see evidence for distinct citation behaviors---a preference for other woman author teams when citing familiar work, and citation equity when citing less familiar work---consistent with practices of resistant knowledge building in other parts of society \cite{Collins2019}.

Our results point toward specific actions that can be taken by individual citers and publishing journals to mitigate citation inequity within physics.
For individual authors, we advocate for thoughtful engagement with the gender make-up of every published reference list, and highlight an increasingly common accountability measure in the form of a citation diversity statement \cite{Zurn2020, Dworkin2020a}.
In particular, our work indicates that special care must be taken when referencing work that lies outside one's immediate sphere of knowledge or scientific network; in these cases, over-citation of man-authored papers can be especially high.
We also show that longer reference lists tend to display less over-citation of man-authored papers, for all citing teams. 
These results suggest an implicit gendered meritocracy whereby man authored papers are viewed as more deserving than woman authored papers when resources (reference list length) are limited. 
While addressing this implicit gendered meritocracy, individual researchers might also choose to cite more papers (when allowed) to broaden engagement with other scientists and work toward citation equity.
Publishing journals, similarly, can reconsider limitations on reference list lengths, recommend the inclusion of citation diversity statements, and work toward more equitable representation in their author pool, with the aim of mitigating citation imbalance in their current and future articles.

\section{Methods}

In this section we will briefly describe our methods for data acquisition, preprocessing, and analysis. 
Our methods draw from and build upon the broad and growing body of literature on gendered citation bias throughout academia \cite{Sugimoto2013, Ghiasi2015, Caplar2017, Dworkin2020, Wang2021, Dion2018, Maliniak2013, Mitchell2013, Fulvio2020, Chatterjee2021}; in particular, we have expanded upon the methods used for a recent study of citation bias in neuroscience journals \cite{Dworkin2020}.

\subsection{Data acquisition}
\subsubsection{Journal selection}
In order to broadly characterize citation behavior within and across the subdisciplines of physics, we selected a list of peer-reviewed journals guided by several criteria. 
First, the journals needed to cover all major subfields of physics, as defined by the breakdown of the \emph{Physical Review} family of journals and the \emph{Web of Science} database categories. 
Second, within each subfield we selected the central journals based on their Eigenfactor score \cite{bergstrom_eigenfactortm_2008} (a measure that rates journals according to incoming citations, weighted by the impact of those citations' journals) as reported by InCites Journal Citation Reports for the year 2018.
Third, we aimed to represent each subfield by an equal number of journals, while allowing the number of papers to differ. This initial search resulted in 7 non-overlapping lists of 5 journals each, for a total of 35 journals.
These journals are listed in Fig. \ref{fig:overview}b.

\subsubsection{Data collection}
We downloaded all papers published in the 35 chosen journals between 1995 and 2020 from the \emph{Web of Science} database.
We then selected all papers classified as original research articles or review articles, and extracted information on author names, reference lists, publication dates, and DOIs.
Each paper's citation behavior was obtained by matching DOIs contained within its reference list to DOIs of papers in our dataset.
Authors' last names were included for all papers; however, for a portion of papers, authors' first names were not included in the database.
First names are necessary for our name-based author gender categorization scheme (see Section \ref{sec:AGC}), so we attempted to find these missing first names in two ways: (i) Searching for author first names using Crossref's API, and (ii) Implementing a name disambiguation scheme (see Section \ref{sec:disambig}) whereby authors' initials and/or other name abbreviations are matched to and replaced by their full first names if possible.

\subsection{Data preprocessing}
\subsubsection{Author name disambiguation} \label{sec:disambig}
To maximize the number of analyzable papers for subsequent author gender categorization by first names, we employed a method of disambiguating authors for whom several versions of their name or initials were available across papers, followed by assignment of the most complete version of each author's name to all papers they authored.
Due to the size of our dataset, we performed our name disambiguation procedure in parallel on isolated subsets of papers, grouped according to their initially-defined subfields shown in Fig S1.
This method involved several steps.
First, we identified all cases for which first and/or last authors' first names consisted only of initials by isolating all authors' first names for each paper, and then flagging the first and last of those which contained only uppercase letters.
Next, for each case of a flagged first name that only contained initials, we gathered all other name instances with the same first/middle initials and the same last name.
If only one unique first/middle full name matched the initials-only entry, or if all distinct full name matches were variants of the same name, we assigned that name to the initials.
However, if multiple unique first/middle full names matched the initials-only entry, we did not assign a name to the initials.
For example, if an entry listed an author as A. A. Griffin, and we found matches under Abby A. Griffin and Abigail A. Griffin, we would replace the A. A. Griffin entry with Abigail A. Griffin. 
If instead we found matches under Abby A. Griffin and Arlene A. Griffin, we would not assign a name to A. A. Griffin.
Initially, our dataset contained 463,538 initials-only first and/or last author name entries. 
Through the steps outlined above, we were able to assign full first names to 88,780 of these entries, for a success rate of $88,780/463,538 \approx 19.15\%$.

We employed a similar strategy to match name variants in order to more accurately determine authorship histories of individual authors.
For every author entry, we first identified sets of corresponding author entries with matching last names and either the same first name or first names determined to be common nicknames of each other (e.g., Abby and Abigail) according to the Secure Open Enterprise Master Patient Index \cite{toth_soempi:_2014}.
If no corresponding matches existed, we retained the author entry name.
If corresponding matches did exist and one match occurred more often, the less common name variant was changed to the more common name variant.
Similarly, if the corresponding matches did not have any conflicting initials, the less common initial variants were changed to the more common initial variants.
A common non-conflicting scenario in this case was that some matches had middle initials and others had no middle initials.
If multiple corresponding matches did have conflicting initials, however, the author entry name was not changed.

\subsubsection{Estimation of publication month}
While the year of publication was available for all 1,067,276 papers, the publication month was not available for 38,423 papers ($\approx 3.6\%$). 
In order to approximate the unknown month $m_i$ for each paper $i$ published in year $y_i$, we considered its lower and upper bounds. 
The lower bound was set to be the month of publication of the most recent paper cited by $i$ if the most recent paper was published in year $y_i$, or January otherwise. 
The upper bound was set to be the month of publication of the first paper to cite $i$ if the first paper was published in year $y_i$, or December otherwise. 
We then approximated $m_i$ as the midpoint between the upper and lower bounds.
To assess the validity of this approach, and to understand the associated uncertainty, we performed the same analysis on the $1,067,276 - 38,423 = 1,028,853$ papers for which the publication month was available. 
In this test case, we found that the average absolute error between the true month and the estimated month was $\approx 2.27$ months. 
In contrast, the average absolute error of naively guessing a publication month between June and July was $\approx 2.98$, indicating that our method provides a reasonable approximation.

\subsubsection{Name-based assignment of author gender categories} \label{sec:AGC}
We assigned ``author gender categories" to papers according to first names of papers' first and last authors.
Gender was assigned to authors with available first names using Gender API, a paid service that includes statistics for approximately 6 million unique first names across 191 countries at the time of writing \cite{genderAPI}.
We assigned the label `man' (`woman') to each author if their first name had a probability $\geq 0.7$ of belonging to someone labeled `man' (`woman') according to our sources \cite{Dion2018}.
Labels are assigned in the Gender API dataset according to a combination of sex assigned to children at birth or chosen by adults later, and gender detected in social media profiles.

Our dataset includes $n=1,067,276$ papers, for a total of $n$ first authors and $n$ last authors. 
Among them, we were able to assign gender labels to both first and last authors in 60\% of papers, to only one of either first or last author in 27\% of papers, and to neither the first nor last author in 13\% of papers. 
Among all first and last authors with unassigned gender labels, 65.4\% were due to publishing using initials.
To ensure that omitting these authors in subsequent analyses would not significantly skew our results, we sought to estimate the gender label distribution among these authors.
As a proxy for this author set, we examined the set of authors for which we uniquely matched initials to first names using our name disambiguation scheme, and successfully assigned author gender labels (see Section \ref{sec:disambig}).
Of these 79,592 authors, 66,255 ($\approx 83.24\%$) were assigned the label `man' by our algorithm and 13,337 ($\approx 16.76\%$) were assigned the label `woman' (see the \emph{Supplementary Information} for details). 
These numbers are consistent with the ratios of assigned gender labels of all first and last authors in our dataset, of which $\approx 87.92\%$ were labeled `man' and $12.07\%$ were labeled `woman.' 
Hence, as far as we can measure, the distribution of genders in unassigned author names is not radically different than that of assigned author names.

We then subdivided papers into author gender categories according to assigned gender labels of first and last authors.
If the first and/or last author's name was assigned the label `woman,' we categorized the paper as W$\vert \vert$W.
To increase statistical power, we included all papers in this category with at least one woman-assigned first/last author, even if the other author could not be assigned a gender label according to our methods.
If the first and last author's name was assigned the label `man,' we categorized the paper as MM. 

We emphasize that, although the phrase ``author gender category" contains the word ``gender," it need not capture the true gender identities of all authors.
Instead, it expresses a statistical correlation: 
By `woman' we mean an author whose name has a probability greater than or equal to 0.70 of belonging to someone identifying as a woman on social media or in federal documents; likewise, by `man,' we mean an author whose name has a probability greater than or equal to 0.70 of belonging to someone identifying as a man on social media or in federal documents.
True gender identity could only be learned through careful manual research of self-attested gender identity, or already known through kinship or conversation, and is not accessible via an automated analysis pipeline like the one used in this paper.
However, the author gender category is nevertheless a notably useful proxy of gender for the purposes of this paper, because it expresses a correlation between name and gender.
Names greatly influence perceptions of gender identity \cite{Pilcher2017}, with well-known implications for a person's perceived merit as a scientist \cite{Moss-Racusin2012}.
These perceptions have marked power to shape citation behavior, irrespective of authors' true gender identities.

\subsubsection{Reference list cleaning procedure}
We pared down each paper's reference list into a ``clean" version suitable for our analysis, which we used for all investigations of citation behavior (unless indicated otherwise).
From each reference list, we removed self-citations, citations to papers not in our dataset, and citations to papers with authors whose names could not be assigned to a gender category.
We chose to remove self-citations from all analyses of citation behavior in order to focus on the influence of perception of other authors' gender category on external citation behavior.
Self-citations were defined conservatively as references to papers for which either the cited first or last author was also the citing first or last author.
We note that in a related study of citation behavior in neuroscience journals, it was shown that including self-citations did not result in meaningful differences in overall over-/under-citation trends \cite{Dworkin2020}.

\subsection{Analysis}
\subsubsection{Probability estimation of author gender category according to paper characteristics}
To quantify over-/under-citation behavior, we first developed a gender-blind null model predicting the probability that papers were written by MM or W$\vert \vert$W author teams according to a set of paper characteristics enumerated below.
Then, over any set of reference lists, we could tally the number of cited papers written by MM or W$\vert \vert$W author teams and compare these quantities to the expected numbers given by the gender-blind null model.
The characteristics we used to predict  MM or W$\vert \vert$W authorship for each paper were (i) month and year of publication, (ii) first and last authors' combined number of papers in the dataset, (iii) total number of authors, (iv) publishing journal, and (v) categorization as a research or review article (Fig. \ref{fig:GAM}a).
Specifically, we fit a generalized additive model (GAM) on the binomial outcome \{MM, W$\vert \vert$W\} with predictive features as defined above.
We fit this model to all papers in our dataset whose authors' names could be assigned gender categories.
To fit the GAM, we utilized the `mgcv' package in R \citep{mgcv}, using penalized thin plate regression splines for estimating smooth terms of features (i), (ii), and (iii) described above.
We note that we used the logarithm of feature (iii) and a winsorized version of feature (ii), capped at 300 (representing the top 0.6\% of papers), to ensure that we fit the GAM successfully.
Univariate thin plate splines were used for the smooth terms, and no interactions between variables were included in the model. 
For each article, the GAM then yields a predicted probability of MM or W$\vert \vert$W\ authorship; we interpret and utilize these probabilities as approximations of the proportion of similar articles (i.e., articles with comparable values of the above characteristics) written by each group. 
These values thus allow us to calculate the proportions of MM and W$\vert \vert$W citations that would be expected if references were drawn in a gender-agnostic manner from pools of characteristic-matched papers.

\subsubsection{Calculation of over-/under-citation}
For a given group of citing papers, we defined its over/under-citation of each author gender category as the percent difference of observed citations from gender-blind expectation.
Over-/under-citation of MM papers, for example, is given by $(o_{MM}-e_{MM})/e_{MM}*100$, where $o_{MM}$ is the (observed) number of citations given to MM papers by the citing papers, and $e_{MM}$ is the expected number of MM citations predicted by the gender-blind model described in the previous section.
More specifically, $e_{MM}$ is computed by summing over the GAM-estimated probabilities that each citation given by the group belongs to the author gender category MM.

\subsubsection{Subfield delineation}
To understand how citation practices vary between disciplines, we grouped journals into defined ``subfields."
The boundaries between subfields were drawn according to a combination of (i) pre-defined journal categories culled from the breakdown of the \emph{Physical Review} family of journals and journal categories defined by \emph{Web of Science}, and (ii) \emph{post-hoc} citation network clustering.
The resultant subfields and their constituent journals are shown in Fig. \ref{fig:overview}b.
See the \emph{Supplementary Information} for further details regarding the citation network clustering that was used to identify subfields.

\section{Results}

\subsection{Time-varying demographics of published papers}

\begin{figure*}
    \centering
    \includegraphics[width = .9\textwidth]{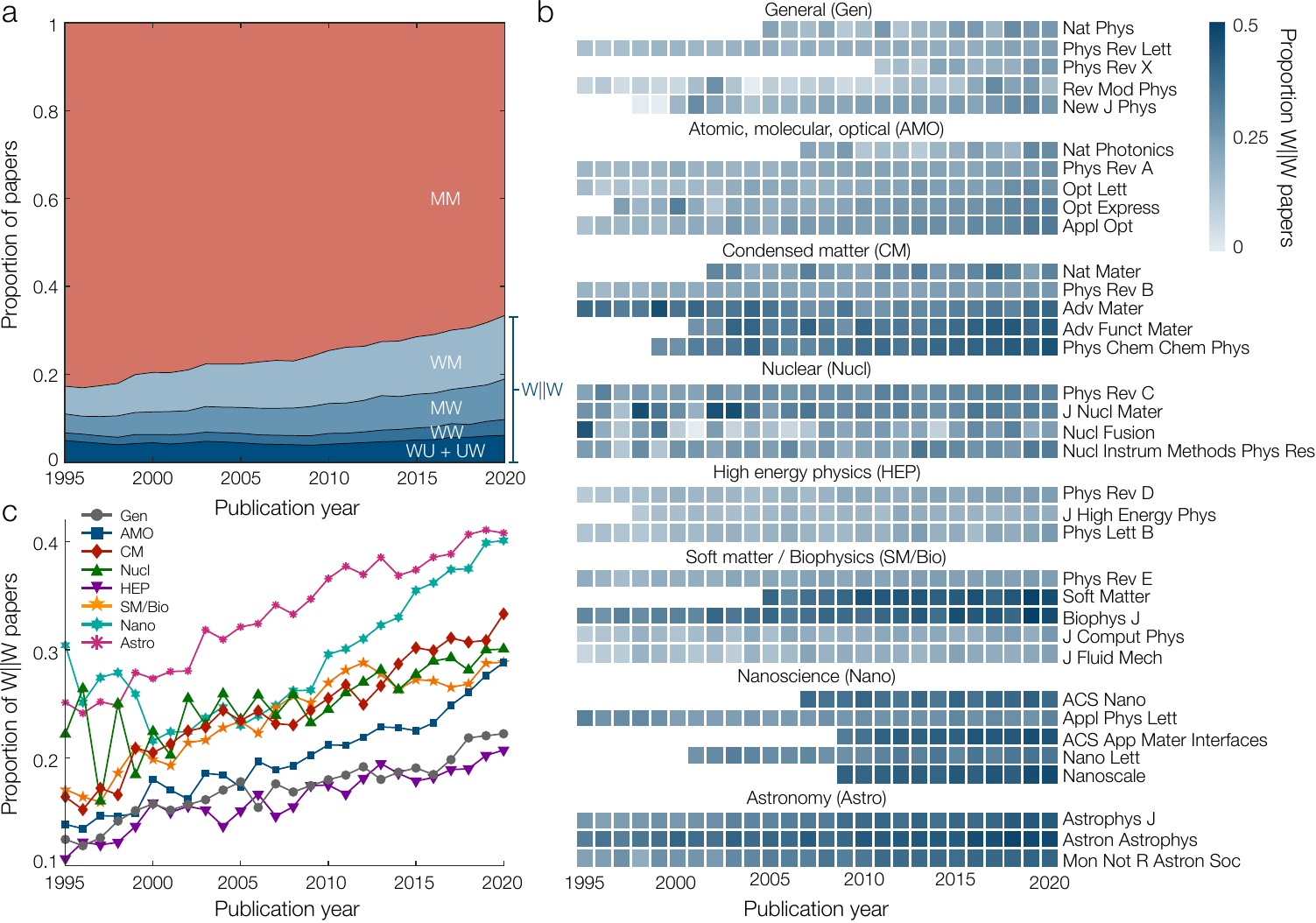}
    \caption{
    \textbf{Time-varying demographics of published papers according to author gender category.}
    (a) Proportions of papers according to author gender category over time.
    The WU + UW sub-category consists of papers in which one of the (first or last) author's names was assigned to the woman gender category, and the other of the (first or last) author's names could not be assigned to a gender category (see \emph{Methods}).
    (b) Proportion of W$\vert \vert$W papers published in all journals in our dataset over time.
    Journals are subdivided according to the subfield definitions used throughout the paper.
    Missing squares occur for journals before their launch year.
    (c) Proportion of W$\vert \vert$W papers published within each subfield over time.
    All proportions shown here are with respect to the per-year sum of MM and W$\vert \vert$W papers (that is, not including papers with authors whose names could not be assigned a gender category).
    }
    \label{fig:overview}
\end{figure*}

We first present a demographic overview of the data set.
Our data consists of approximately 1.07 million papers published between the years 1995 and 2020 in 35 representative and central journals (as measured by their 2018 Eigenfactor score) across a range of 8 physics subfields.
Of these papers, 668,690 could be identified according to their name-based author gender category (see the \emph{Methods} for details).
The proportion of papers in each author gender category are shown in Fig.~\ref{fig:overview}a as a function of publication year. 
Although the proportion of woman-authored (W$\vert \vert$W) papers has increased from 17\% in 1995 to 33\% in 2020, it nevertheless represents a small proportion of the total number of MM and W$\vert \vert$W papers, in agreement with other studies \cite{Holman2018}.
Moreover, the proportion of papers authored by first and last authors with names assigned to women (WW) remains significantly lower, growing from 1.7\% in 1995 to only 3.6\% in 2020.

Although the proportion of W$\vert \vert$W papers increases in all subfields, individual journals vary significantly in that proportion and in its rate of change (Fig. \ref{fig:overview}b-c). 
The journals grouped into the general physics subfield---which includes many of the highest impact journals in physics---and the high energy physics subfield collectively contain the lowest proportions of W$\vert \vert$W papers. 
This pattern remains consistent over time. 
Conversely, journals grouped into the astronomy and astrophysics subfield generally contain the highest proportion of W$\vert \vert$W papers over time, followed by the relatively young journals grouped into the nanoscience subfield (four of which were launched during the time period we analyze).
Averaged over years, journals with the lowest fraction of W$\vert \vert$W papers are \emph{Reviews of Modern Physics} (0.11), \emph{Journal of High Energy Physics} (0.15), and \emph{Journal of Fluid Mechanics} (0.15).
These three journals also have the lowest fraction of W$\vert \vert$W papers published in 2020 among all journals, indicating that this trend is not merely due to their older age and the general increase of woman authors over time.
Journals with the highest fraction of year-averaged W$\vert \vert$W papers are \emph{Nanoscale} (0.41), \emph{Soft Matter} (0.40), and \emph{ACS Applied Materials \& Interfaces} (0.40).

\subsection{Citation imbalance exists \& varies by citing venue}

We next examine the citation behavior of the papers in our dataset, and find that citation behavior is imbalanced with respect to the author gender category of the cited papers.
We consider only the citation behavior of papers published in 2009 or later, since these papers are more likely than those published earlier to cite other papers in our (1995 -- 2020) dataset.
To increase statistical power, the citation behavior reported in this section is aggregated over all papers published in 2009 or later, even those that could not be assigned to an author gender category.
We define over-/under-citation of an author gender category as the percent difference between the number of citations that papers in the category actually receive and their expected number of received citations according to a gender-blind model (see \emph{Methods}).

\begin{figure*}
    \centering
    \includegraphics[width = .8\textwidth]{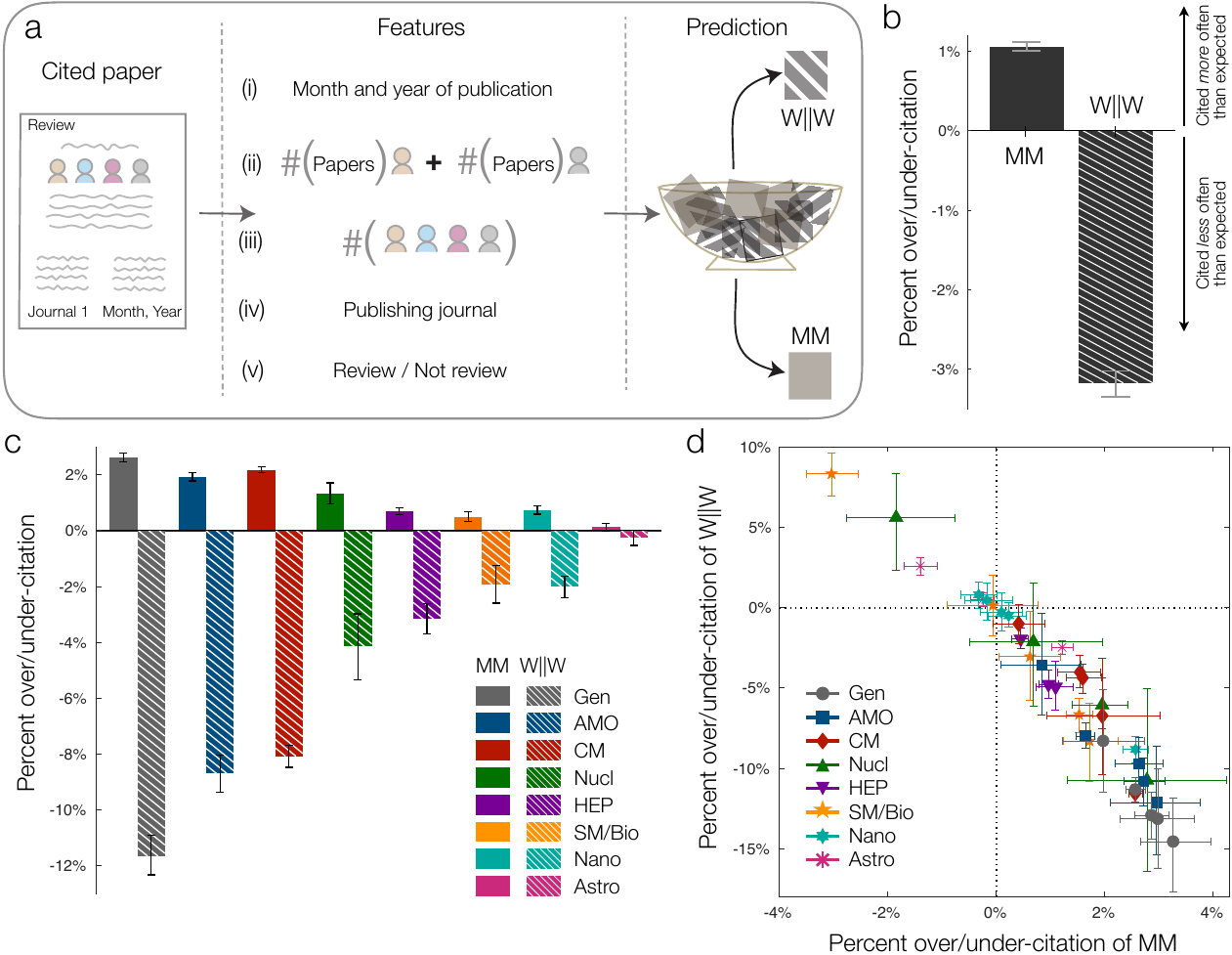}
    \caption{
    \textbf{Over-/under-citation of physics papers is imbalanced with respect to author gender category.}
    (a) Illustration of the statistical model used to estimate over-/under-citation of MM or W$\vert \vert$W papers according to paper characteristics.
    (b) Over-citation of MM papers and under-citation of W$\vert \vert$W papers is exhibited in aggregate over all citing papers in our data set published between 2009 and 2020.
    Over-/under-citation varies when the citing papers are grouped according to their
    (c) subfield and 
    (d) journal.
    In panel d, each marker reflects an individual journal colored according to subfield.
    In all panels, the reported over-/under-citation values utilize the reference lists of all relevant citing papers, including those of unknown author gender category, to increase statistical power. Error bars representing the 95\% CI of each over-/under-citation calculation were computed via 500 bootstrap resampling iterations.
    }
    \label{fig:GAM}
\end{figure*}

In general, MM papers are cited \emph{more} often than expected by approximately 1.06\%, and W$\vert \vert$W papers are cited \emph{less} often than expected by approximately 3.17\% (Fig. \ref{fig:GAM}b). 
These results constitute an overall gender citation gap of roughly 4.23\%. 
This gap varies significantly across subfields (Fig. \ref{fig:GAM}c) and journals within those subfields (Fig. \ref{fig:GAM}d). 
Papers published in journals grouped into the general physics subfield collectively show the widest citation gap, whereas papers in journals grouped into the astronomy and astrophysics subfield collectively show the narrowest citation gap. 

As shown in finer detail in Fig. \ref{fig:GAM}d, citation behavior at the journal level varies more widely still. 
Here, each point in the space of MM over-/under-citation versus W$\vert \vert$W over-/under-citation shows the collective citation imbalance of papers grouped according to their publishing journal.
The data follow an approximately linear trend with negative slope, indicating the correlation between over-citation of MM papers and under-citation of W$\vert \vert$W papers.
Journals that lie more deeply in the lower right quadrant host papers that collectively exhibit greater citation preference for MM papers, whereas journals that lie in the upper left quadrant host papers that collectively exhibit greater citation preference for W$\vert \vert$W papers. 
Most journals are located in the lower right quadrant, indicating their preference toward MM over-citation, and only three journals (\emph{Journal of Nuclear Materials}, \emph{Astronomy \& Astrophysics}, and \emph{Soft Matter}) show a statistically significant preference for citing W$\vert \vert$W papers. 

\begin{figure*}
    \centering
    \includegraphics[width = .8\textwidth]{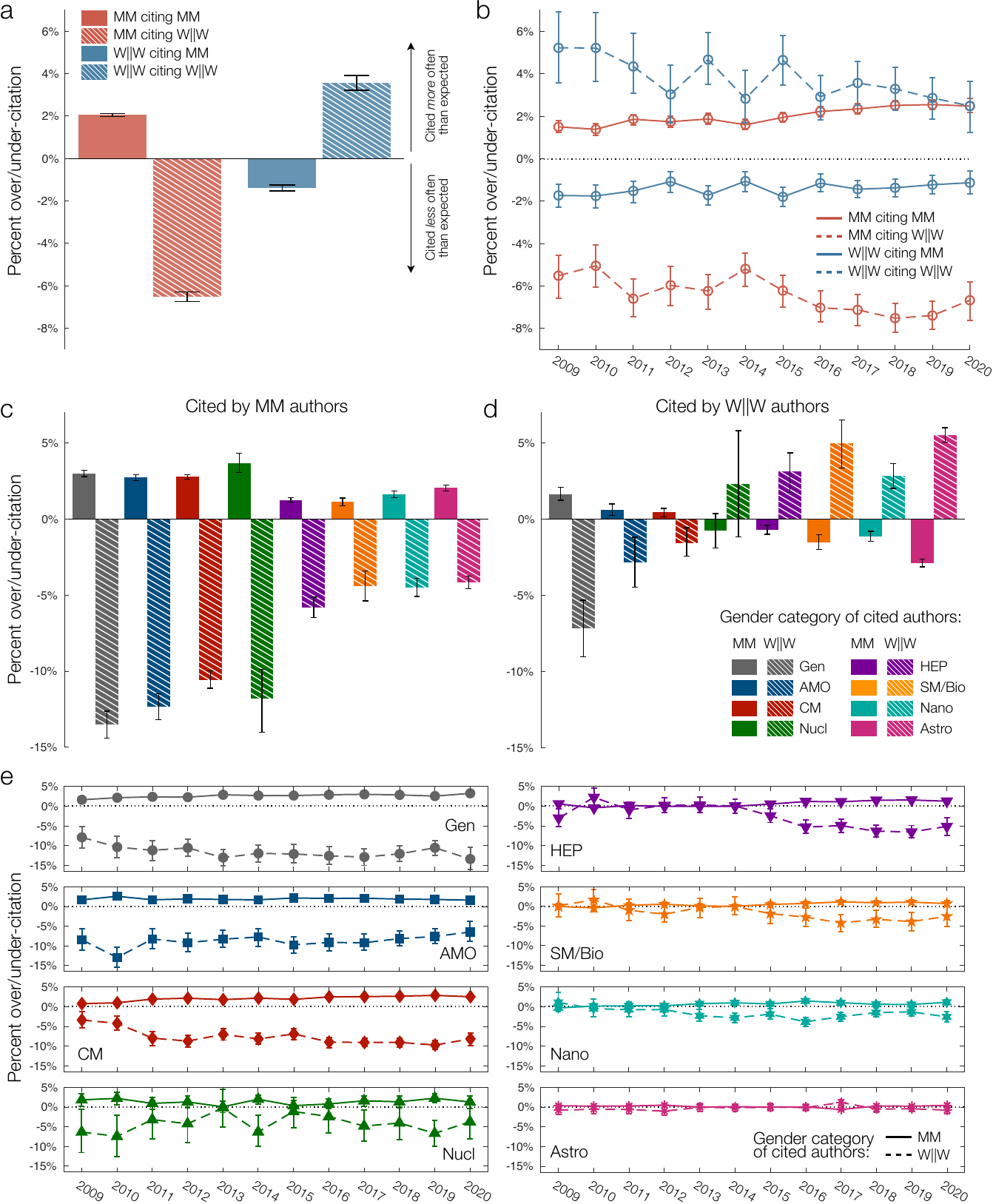}
    \caption{
    \textbf{Citation behavior varies across time and according to citing author gender category.}
    (a) Over-/under-citation of MM and W$\vert \vert$W papers, calculated separately for MM and W$\vert \vert$W citing teams published between 2009 and 2020.
    Each team exhibits citation preference toward their own author gender category.
    (b) Citation behavior of each citing author gender category varies over time. MM citing teams exhibit a higher citation preference toward other MM papers and a lower citation preference to W$\vert \vert$W papers, resulting in a citation gap that is widening over time.
    W$\vert \vert$W citing teams exhibit lower citation preference toward other W$\vert \vert$W papers, and that preference decreases over time.
    (c-d) Citation imbalance exhibited by MM teams and W$\vert \vert$W teams varies according to subfield.
    In both types of teams, papers in the general physics category exhibit the highest over-citation in favor of MM papers.
    (e) Overall citation imbalance within subfields is relatively stable over time.
    Values reported in this panel incorporate the reference lists of all relevant papers, including those of unknown author gender category, to increase statistical power.
    In all panels, error bars representing the 95\% CI of each over-/under-citation calculation were computed via 500 bootstrap resampling iterations.
    }
    \label{fig:GAMtime}
\end{figure*}

\subsection{Citation imbalance varies by citing actor}

We further separate citation behavior according to the author gender category of citing teams, and find that MM papers and W$\vert \vert$W papers differ in their citation behavior: MM papers tend to exhibit higher citation preference toward other MM papers and lower citation preference toward W$\vert \vert$W papers, whereas W$\vert \vert$W papers tend to exhibit higher citation preference toward other W$\vert \vert$W papers and lower citation preference toward MM papers.
The extent and existence of these citation preferences vary across subfield and across journal.
MM citing papers published in 2009 or later over-cite other MM papers by 2.05\%, and under-cite W$\vert \vert$W papers by 6.53\%, for a gender citation gap of approximately 8.58\% (Fig. \ref{fig:GAMtime}a).
By contrast, W$\vert \vert$W citing papers published in 2009 or later over-cite other W$\vert \vert$W papers by 3.56\%, and under-cite MM papers by 1.38\%, for a citation preference in favor of other W$\vert \vert$W papers of approximately 4.94\% (Fig. \ref{fig:GAMtime}a).

These citation behaviors vary widely across subfields (Figs. \ref{fig:GAMtime}c-d, S4).
MM papers in the general physics subfield show the highest citation gap, 16.47\%, in favor of other MM papers.
MM papers in the nuclear and atomic/molecular/optical subfields show the next highest citation gaps, each above 15\%.
By contrast, W$\vert \vert$W papers in the astronomy/astrophysics and soft matter/biophysics subfields exhibit a citation preference toward other W$\vert \vert$W papers, resulting in citation gaps in favor of other W$\vert \vert$W papers of 8.42\% and 6.53\%, respectively.
Interestingly, citation preference in favor of other W$\vert \vert$W papers does not exist for W$\vert \vert$W citing papers in the general physics, atomic/molecular/optical, and condensed matter subfields.
Rather, papers in these subfields show under-citation of other W$\vert \vert$W papers and over-citation of MM papers.
W$\vert \vert$W papers grouped into the general physics subfield show the greatest citation gap (8.8\%) in favor of MM papers.

\subsection{Stable and growing trends in citation imbalance over time}

Inspection of citation behavior as a function of time reveals that, in general, the citation gap between over-citation of MM papers and under-citation of W$\vert \vert$W papers has remained relatively stable between the years of 2009 and 2020, and even slightly grown (Fig. \ref{fig:GAMtime}b,e).
The difference in citation preference between MM and W$\vert \vert$W citing teams, explored in the previous section, also persists over time.
However, time trends in these citation preferences vary significantly for MM versus W$\vert \vert$W citing teams.
Trends also differ across the subfield and publishing journal of citing papers; the overall picture, however, is one of stagnant or worsening citation gaps over time, despite the relative growth of papers authored by W$\vert \vert$W teams.

The time evolution of citation imbalance for all papers in our dataset from 2009 to 2020, grouped according to citing author gender and year of publication, is shown in Fig. \ref{fig:GAMtime}b.
The citation gap in favor of MM papers exhibited by MM citers is larger in 2020 than in 2009 (Fig. \ref{fig:GAMtime}b, red lines), due to an overall increase in MM over-citation and W$\vert \vert$W under-citation over time.
That this gap persists in the face of a growing fraction of published W$\vert \vert$W papers over the same time period (Fig. \ref{fig:overview}a) is notable.
The fraction of citations actually given by MM citers to MM (W$\vert \vert$W) papers over time maintains a steady positive (negative) gap with respect to the fraction expected to be given according to our model (Fig. S5a). 
By contrast, W$\vert \vert$W citers show a consistent citation preference in favor of other W$\vert \vert$W papers with a markedly different trend over time (Fig. \ref{fig:GAMtime}b, blue lines).
For these citers, the citation gap in favor of other W$\vert \vert$W papers decreases over time, in contrast to the increasing citation gap in favor of other MM papers shown by MM citers.
The fraction of citations given by W$\vert \vert$W citers to MM (W$\vert \vert$W) papers over time shows a narrowing negative (positive) gap with respect to the fraction expected to be given according to our model (Fig. S5b).
The data indicate an approach toward citation parity over time by W$\vert \vert$W citers.

When grouped according to subfield, citation behavior shows a variety of different trends over time (Figs. \ref{fig:GAMtime}e, S6, S7).
Citation gaps between MM authored papers and W$\vert \vert$W authored papers vary in magnitude according to subfield, but are relatively stagnant over time in many cases (within error bars).
Two subfields that notably do not exhibit this behavior are the general physics and condensed matter subfields, for which the citation gap increases between 2009 and 2020, to 16.67\% and 10.7\% in 2020, respectively.

\begin{figure*}
    \centering
    \includegraphics[width = .8\textwidth]{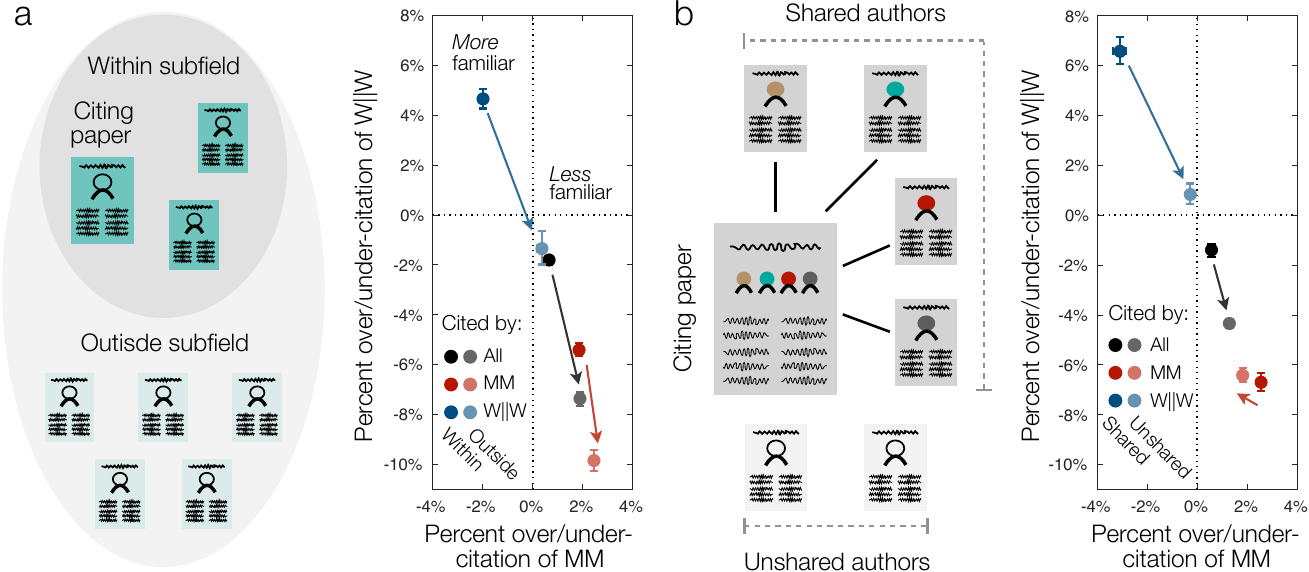}
    \caption{
    \textbf{Citation behavior varies according to proxies for familiarity with cited work.}
    Proxies are 
    (a) whether the publishing journals of cited and citing papers fall within the same subfield, and 
    (b) whether the cited and citing papers have any shared authors.
    In each panel, a schematic of the proxy for familiarity is shown on the left, and citation behavior according to that proxy is shown on the right.
    Markers show the citation behavior of (black) all papers written between 2009 and 2020, including those with unassigned author gender categories, (red) the subset of papers in the MM author gender category, and (blue) the subset of papers in the W$\vert \vert$W author gender category.
    Arrows point from over-/under-citation when considering only ``familiar" citations to over-/under-citation when considering only ``unfamiliar" citations.
    Overall, over-citation of MM papers and under-citation of W$\vert \vert$W papers is more marked for unfamiliar citations, although this effect is due to differing behaviors of MM and W$\vert \vert$W citing teams.
    In both panels, error bars representing the 95\% CI of each over-/under-citation calculation were found via 500 bootstrap resampling iterations.
    }
    \label{fig:subfield}
\end{figure*}

\subsection{Citation imbalance varies according to its form}

The citation behaviors exhibited by MM and W$\vert \vert$W teams vary significantly in magnitude according to the form or type of each citation; more specifically, whether the citation references work that authors are more or less likely to be familiar with.
We define proxies for familiarity between citing paper and cited paper in two ways: (i) whether they are published in journals that fall within the same subfield (Fig. \ref{fig:subfield}a, left), and (ii) whether they are written by at least one shared author (Fig. \ref{fig:subfield}b, left, and \emph{Supplementary Information}).
Reference lists of citing papers contain a range of reference proportions that are deemed familiar according to these definitions (Fig. S8).
We additionally define subfield boundaries by clustering the citation network between journals (see \emph{Supplementary Information}), resulting in subfields distinct from those shown here, and find that our results are robust across subfield definition (Fig. S9).

With both proxies for familiarity, we find similar results regarding the total citation gap between MM and W$\vert \vert$W authored papers: For the set of unfamiliar citations, the citation gap is larger, with greater over-citation of MM papers and greater under-citation of W$\vert \vert$W papers (Fig. \ref{fig:subfield}a-b, black symbols).
Out-of-subfield MM papers are over-cited by 1.90\%, and out-of-subfield W$\vert \vert$W papers are under-cited by -7.36\%, while these rates are 0.66\% and -1.80\% for within-subfield MM and W$\vert \vert$W papers, respectively.
Unshared-author MM papers are over-cited by 1.29\%, and unshared-author W$\vert \vert$W papers are under-cited by -4.34\%, while these rates are 0.57\% and -1.39\% for shared-author MM and W$\vert \vert$W papers, respectively.
For reference, we note that these values for all forms of citation, independent of familiarity, are MM over-citation of 1.06\% and W$\vert \vert$W under-citation of 3.17\% (Fig. \ref{fig:GAM}b). 

The suppression of the ``familiar citation gap" and the enhancement of the ``unfamiliar citation gap" together arise from the cumulative effects of two very different citation behaviors according to citing author gender category.
Across familiar citations, both MM (Fig. \ref{fig:subfield}a-b, red symbols) and W$\vert \vert$W (Fig. \ref{fig:subfield}a-b, blue symbols) citing teams show enhanced citation preference for their respective author gender categories. 
That is, MM teams preferably cite MM papers, while W$\vert \vert$W teams preferably cite W$\vert \vert$W papers.
We note that these competing effects may be partially explained by an overall homophilic enhancement in the local co-authorship network around each paper, revealed by higher man-author overrepresentation in co-authorship networks around MM papers and lower man-author overrepresentation in co-authorship networks around W$\vert \vert$W papers (see \emph{Methods}, Fig. S10).

Across unfamiliar citations, however, MM and W$\vert \vert$W teams differ in their citation behavior.
For W$\vert \vert$W teams, citation preference for W$\vert \vert$W papers is approximately erased.
By contrast, for MM teams, citation preference for MM papers is not erased.
Instead, this preference is enhanced for out-of-subfield citations, and slightly reduced but still significant for unshared-author citations.
Thus, we find that W$\vert \vert$W teams preferably cite familiar W$\vert \vert$W papers and cite unfamiliar W$\vert \vert$W and MM papers approximately equitably. 
MM teams over-cite familiar MM papers and under-cite familiar W$\vert \vert$W papers, a trend that is even more pronounced for unfamiliar papers.
The overall result is a smaller familiar citation gap, due to the competing citation behaviors of W$\vert \vert$W and MM citing teams, and a larger unfamiliar citation gap in favor of MM papers, due to the MM citing papers' persistent citation preference for other MM papers even when citing unfamiliar references.

Note that, although citation imbalance across familiar citations is approximately equal and opposite for W$\vert \vert$W and MM citing teams (for both definitions of familiarity), it is not distributed equally across W$\vert \vert$W and MM cited papers.
Rather, W$\vert \vert$W teams exhibit a citation preference for W$\vert \vert$W papers (vertical coordinate of the dark blue markers in Figs. \ref{fig:subfield}a,b) that is about double that of MM teams for MM papers (horizontal coordinate of the dark red markers in Figs. \ref{fig:subfield}a,b).
By contrast, MM teams under-cite W$\vert \vert$W papers (vertical coordinate of the dark red markers in Figs. \ref{fig:subfield}a,b) at a rate that is about double the under-citation of MM papers by W$\vert \vert$W teams (horizontal coordinate of the dark blue markers in Figs. \ref{fig:subfield}a,b).

\subsection{Additional correlates of citation imbalance}
\begin{figure*}
    \centering
    \includegraphics[width = \textwidth]{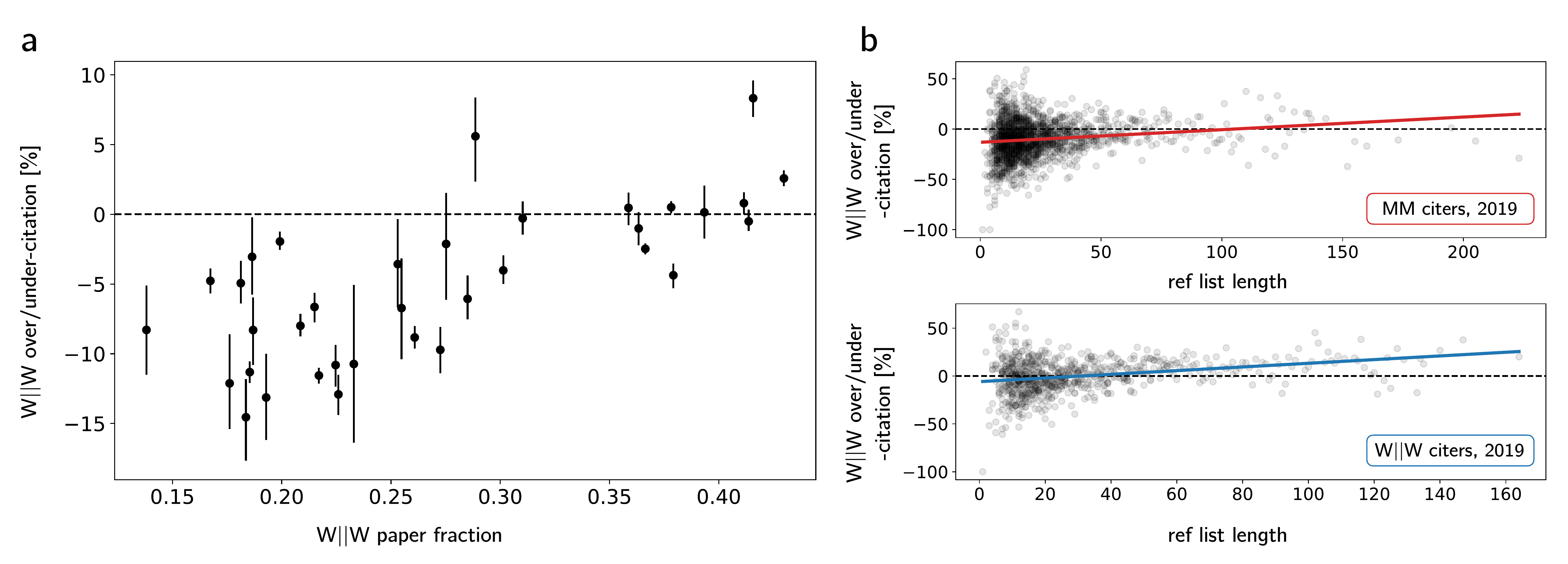}
    \caption{
    \textbf{Additional correlates of citation behavior.}
    (a) Journals that published a higher proportion of W$\vert \vert$W papers between 2009 and 2020 generally exhibited higher citation of W$\vert \vert$W papers.
    Each marker represents a journal; error bars representing the 95\% CI of each over-/under-citation value (calculated across all citing papers published in each journal between 2009 and 2020) were found via 500 bootstrap resampling iterations.
    Proportions of W$\vert \vert$W papers are reported with respect to total MM and W$\vert \vert$W papers published in each journal between 2009 and 2020.
    (b) Papers with longer reference lists tend to exhibit higher citation of W$\vert \vert$W papers.
    The panels show that this trend is consistent for MM papers published in 2019 (top) and W$\vert \vert$W papers published in 2019 (bottom).
    Each data point shows over-/under-citation aggregated over paper subsets of maximum size 10 at each reference list length (see \emph{Supplementary Information} for details).
    Linear fits to the data via ordinary least-squares are also shown, expressed approximately by $y=0.126x-13.095$ (top) and $y=0.192x-5.911$ (bottom). 
    }
    \label{fig:discussion}
\end{figure*}

Our dataset reveals additional correlations between citation behavior and citing actor, venue, and form, each of which might be useful considerations for developing individual and institutional strategies to mitigate citation imbalance in the future.
In particular, we find that citation behavior varies on the journal level according to the relative proportion of W$\vert \vert$W published papers, and on the paper level according to the length of the reference list. 
For the former, we note that the 35 journals investigated in this paper generally show an increasing time-aggregated citation preference for W$\vert \vert$W papers with an increasing time-aggregated fraction of W$\vert \vert$W papers published (Fig. \ref{fig:discussion}a). 
Correspondingly, the time-aggregated citation preference for MM papers decreases with an increasing time-aggregated fraction of W$\vert \vert$W papers published (Fig. S11). 
These results are not unexpected given the tendency of W$\vert \vert$W papers to cite other W$\vert \vert$W papers to a greater extent than MM citing papers do (Fig. \ref{fig:GAMtime}a-d). 
The data demonstrate that representation within journal authorship pools is a meaningful correlate of citation imbalance.

For individual citing papers, we also find a collective trend whereby papers with longer reference lists tend to exhibit increased citation preference for W$\vert \vert$W papers (see \emph{Supplementary Information}, Fig. S12).
This trend holds independently for both MM and W$\vert \vert$W citing teams, despite the fact that W$\vert \vert$W papers in our dataset contain longer reference lists in comparison with MM papers (Fig. S13). 
Notably, the trend also remains stable over time, despite the fact that reference list length within our dataset generally increases over time (Fig. S14).
Fig. \ref{fig:discussion}b demonstrates this phenomenon for papers published in 2019. 
Papers are grouped according to author gender category and reference list length, and linear fits to the data for each author gender category show the upward trend of W$\vert \vert$W citation preference with reference list length.
The slopes of these linear fits, averaged over the years between 2009 and 2020, suggest that reference lists gain approximately 1.5\% in W$\vert \vert$W citation preference when they add 10 citations (Fig. S15).
Because the coefficient of determination of each linear fit is low (Fig. S16), the relation should be considered a weak but statistically significant collective effect rather than an exact measure of correlation for any particular paper subset.
The conceptual significance of this effect is notable when considered in light of the global rate of MM over-citation (1.06\%) and W$\vert \vert$W under-citation (-3.17\%) for all (2009-2020) papers (Fig. \ref{fig:GAM}b).

\section{Discussion}

Gender disparities exist throughout the discipline, performance, and processes of science. 
While visible markers of these disparities have been detectable for decades (or rather, centuries), other manifestations of imbalance are more subtle, requiring the use of tools and technologies in the science of science. 
Employing such tools, here we focused on evaluating the extent and drivers of gender imbalance in the reference lists of physics papers published across a broad range of 35 physics journals over the last 25 years. 
Using automated name-based gender prediction and a model for citation behavior that estimates over/under-citation while controlling for paper characteristics, we found that papers within the man author category have been globally over-cited, and papers within the woman author category have been globally under-cited.
We additionally found that this citation imbalance depends heavily on citing author gender category, the publishing journal of citing papers, proxies for familiarity of citing authors with cited papers, representation in journal authorship pools, and reference list length. 
Our results underscore the complexity with which gender disparities manifest in scientific publishing, and identify correlates that could inform future actions by individuals, journals, and collectives. 

\subsection{Homophilic behavior and its drivers}
Throughout our analyses, we observed homophilic citation behavior for both MM and W$\vert \vert$W teams, with MM citing teams tending to devote more of their reference lists to papers published by other MM teams than predicted by our model, and W$\vert \vert$W citing teams tending to devote more of their reference lists to papers published by other W$\vert \vert$W teams than predicted by our model. 
For MM citing teams, this homophilic behavior was persistent over time, pronounced across subfields, and present for both more familiar and less familiar types of citations.
For W$\vert \vert$W citing teams, homophilic behavior lessened in magnitude over time, varied according to subfield (with some subfields not showing homophily), and was significant only for more familiar citations.

The presence of gendered homophilic citation behavior throughout our dataset is not surprising; indeed, homophily along numerous dimensions of similarity is a well-studied and prominent driver of human activity and social system development \cite{McPherson2001}.
In academia, gender homophily influences patterns of co-authorship and collaboration \cite{Boschini2007, Jadidi2018, Wang2019a, Holman2019, Kwiek2021}, citations \cite{Ghiasi2018}, invitations to give colloquia and other talks \cite{Isbell2012, Debarre2018, Nittrouer2018}, networking \cite{Belle2014, Atzmueller2018}, the nomination of Nobel laureates \cite{Gallotti2019}, and peer-reviewer selection \cite{Helmer2017}.
We note that several of these studies imply that gendered homophilic behavior seems unlikely to be an effective driver of gender equity.
A study of gender homophily within primatologists \cite{Isbell2012} found significant homophily among men in invited symposium talks even though women make up the majority of that field.
Homophily has also been shown to increase with increasing representation of women in other contexts \cite{Boschini2007, Wang2019a}.
Moreover, a study of gender homophily in the peer-review process \cite{Helmer2017} found different patterns of homophily for men versus women, with men engaging in widespread homophilic behavior, and the homophilic behavior of women due to only a few highly homophilic women editors.
Authors of that paper stressed that such a top-down approach toward engaging more women in the peer-review process, dominated by a handful of actors, was less likely to remain stable and effective over time than a more bottom-up approach adopted by a range of actors.

We stress that, although homophilic behaviors of majority and minority populations in any context (here, men and women in physics) might seem similar on the level of quantitative analysis, they need not be driven by similar factors.
Members of a minority population may choose to take part in ``activist choice homophily,'' built on the perception of shared structural barriers and a desire to help overcome those barriers \cite{Greenberg2017}.
The driver of activist choice homophily is notably distinct from other typical drivers of homophily based solely on similarity in perceived attributes \cite{McPherson2001}.
In this paper, the homophilic practices of women in physics are acts of citing other women and thereby permanently embedding them into the scientific canon.
These homophilic practices, if driven by activist choice, could thus be argued to fall under the category of resistant knowledge building and epistemic resistance \cite{Collins2019} against a scientific narrative dominated by men, and especially white men.
In general, our results point toward the need for all citing actors within physics to critically engage with their reference lists, consider the historical and contemporary imbalances they seek to address through their citation practices, and act accordingly.

\subsection{Toward citation equity}

Gender disparities in the citation practices of physics are not surprising. 
Such disparities have been reported in fields as diverse as astronomy \cite{Caplar2017}, economics \cite{Dion2018}, neuroscience \cite{Dworkin2020}, cognitive neuroscience \cite{Fulvio2020}, medicine \cite{Chatterjee2021}, communications \cite{Wang2021}, and international relations \cite{Maliniak2013}. 
The responses of these fields can serve as an example for the physics community as it grapples with its own internal culture. 
Such responses have spanned from individual actions to journal stances and editorial policies. 
At the individual level, scholars have taken it upon themselves to add a citation diversity statement (CDS) to their papers, just before the reference list, stating (i) the importance of citation diversity, (ii) the percentage breakdown (or other diversity indicators) of citations in the paper, (iii) the method by which percentages (or other indicators) were assessed and its limitations, and (iv) a commitment to improving equitable practices in science \cite{Zurn2020}. 
In a retrospective analysis, papers published with a CDS have been found to display citation practices in line with expected rates, without any statistically detectable gender disparities \cite{oudyk2021cleanbib}. 
This finding is notable, and underscores the efficacy of the CDS as an individually-led tool for the mitigation of disparities.  

At the journal level, responses have ranged from statements of support (e.g., from \emph{Nature Reviews Physics} and others \cite{Budrikis2020,natneuroeditorial2020,braineditorial2020}), commentaries \cite{Fairhall2020}, and essays \cite{davies2021promoting,pierce2020women,Llorens2021}, to recommendations and formal policies. 
For example, the \emph{Journal of Cognitive Neuroscience} provides a section in their Author Guidelines regarding citation disparity, supports the use of a CDS, and offers authors a \href{https://postlab.psych.wisc.edu/gcbialyzer/} {\color{blue}journal-specific tool} \cite{Fulvio2020} to estimate the gender citation balance index. 
In a retrospective analysis of papers published in the year following these guideline changes, the editors report a clear reduction of gender citation imbalances \cite{postle2021one}. 
The Biomedical Engineering Society, which publishes four journals, also supports the use of a CDS, and wrote and published a call to action \cite{Rowson2021}, while pointing authors to the following publicly available \href{https://zenodo.org/record/4104748#.X784zc1KiUm }{\color{blue}journal-general tool} \cite{dale_zhou_2020_4104748}. 
Even more broadly, Cell Press---which publishes over 50 journals---now uses an inclusion and diversity \href{https://els-jbs-prod-cdn.jbs.elsevierhealth.com/pb/assets/raw/shared/forms/IandDstatement_form.pdf}{\color{blue}form} to gather information on how the values of diversity and inclusion were supported and upheld throughout the performance of the science and development of the paper under consideration for publication \cite{Sweet2021}. 

These positive actions underscore the advances that can be attained when individual scholars embrace the change indicated by bias research \cite{raymond2013most}, an area of scholarly inquiry that faces a discrimination of its own; gender-bias research is generically underappreciated in academia \cite{cislak2018bias}, and viewed less favorably by men, especially men faculty in STEM \cite{handley2015quality}, which is an unfortunate irony given that bias tends to be preferentially perpetuated by those who think it is not happening \cite{begeny2020some}. 
Foregrounding action-based thinking, our study provides context for existing individual- and journal-led efforts, and new directions for mitigation strategies. 
At the individual level, our work underscores the critical need for self-education, particularly focused on the work being published in general physics journals and in fields adjacent to or far from the author's primary area of expertise. 
This need is motivated by our observations that gender imbalances in citations are greatest within citing articles in general physics journals and when cited articles belong to less familiar fields. 

At the journal level, our work motivates future efforts to test the hypothesis that increasing the number of women authors that a journal publishes would lead to a decrease in that journal's gender imbalance in citations. 
This causal hypothesis to be tested derives from our non-causal observation that journals publishing a higher fraction of papers authored by women tend to show less under-citation of women. 
Increasing the proportion of women authors should be relatively easy for journals like \emph{Reviews of Modern Physics}, which can solicit new reviews on topics from under-represented authors; such a step would be particularly helpful, as \emph{Reviews of Modern Physics} currently has the lowest fraction of $W||W$ papers of all journals examined in our study (0.11). 
Any effort to increase the number of women authors in a journal would also serve to combat (i) the higher writing standards for women than for men \cite{hengel2020publishing}, and (ii) the tendency for submitted work to be perceived as better when it is associated with a man's name compared to a woman's name \cite{krawczyk2016authors,Knobloch-Westerwick2013}. 
Our results also suggest the possibility that removing length limits on reference lists may be beneficial to gender equality, as papers with longer reference lists tended to display less disparity. 
However, we note that this recommendation does not condone the implicit notion that man-authored papers are of greater value and more deserving when resources (reference entries) are limited.

\subsection{Expanding the scope of equitable citation practice}

Although the term ``citation'' in academia typically refers to an entry in a reference list, citations take many forms and therefore conversations about citation equity can be fruitfully expanded. Whether it is a formal reference or an informal mention, in spoken or written communication, a citation is a way of calling forward and raising up specific scholars or instances of scholarship \cite{ahmed2013points, ahmed2017living, ahmed2019use, mott2017citation}. Given the multiple ways in which citations can occur, issues of citation (in)equity are likely more ubiquitous than expected. 

Consider the scientist’s writing life. Whether writing articles, reviews, or grant proposals, we can evaluate the gender balance of our reference lists, but we can also start earlier and go deeper. Who jumps to mind (and to pen, or whiteboard) as we generate ideas for a paper? Who is there along the way as we develop the methods and start interpreting the results? Where do those names ultimately show up in the paper—is it in the introduction, the methods section, the results, or the discussion? Where does a greater (or lesser) diversity of authors appear and what does that reveal about how we conceptualized and contextualized the project? Are citations to women prominent, substantial, and critical to the central questions, tools, and interpretive frames of our work? Or are they appended as supplementary or even as afterthoughts? 

Consider, too, the scientist’s speaking life. Whether teaching, giving talks, or informally conversing with students and colleagues, we can evaluate the gender balance of our spoken references. Who do we teach as central figures in the history of science? Who do we recommend to students and colleagues as scholars of idle interest or scholars making an important contribution to the topic under discussion? Who do we explicitly mention in our talks, write into our slides, or acknowledge as thought collaborators? Whose work do we signal boost through invitations to speak, to co-author, to co-investigate, or to write a review or perspective piece for our journal? Who do we retweet in the most mundane and most meaningful of ways? And are women scholars equitably represented in these moments?   

We look forward to future research along all these vectors, and we invite individual and collective reflection not only on how citation imbalances, obdurate homophily, and other dynamics explored in this paper might appear beyond our reference lists but also on how they might be addressed \cite{massen2017sharing, vasarhelyi2021gender, atir2018how}. 

\subsection{Methodological considerations}
While our study expands the frontiers of our understanding of gendered practices in physics, several limitations motivate future work. In particular we underscore the fact that under-attribution, bias, and discrimination are faced by individuals of many---and intersecting---identities. Future work would do well to unpack citation practices not just along the (cisgender) man/woman binary, but also along other dimensions of difference, including trans and/or nonbinary status, race, ethnicity, class, sexual orientation, and disability. For example, prior work in several areas of science provides evidence for under-attribution according to race and ethnicity; scholars of color are broadly undercited by their academic communities \cite{Bertolero2020,chakravartty2018communication}. These effects compound upon known intersecting gender and class inequalities in hiring at all levels in the academy \cite{Sheltzer2014,clauset2015systematic,way2019productivity}; at the faculty level, the prestige of a doctoral student's PhD-granting institution predicts their placement as a tenure-track faculty member, women generally place worse than men, and increased institutional prestige leads to increased faculty production, better faculty placement, and a more influential position within the discipline \cite{clauset2015systematic,way2019productivity}. Furthermore, sexual orientation is a proven axis of inequality in STEM from classrooms to careers \cite{cech2011navigating,gibney2019discrimination}; lesbian, gay, bisexual, transgender, and queer (LGBTQ) STEM professionals are more likely to experience career limitations, harassment, and professional devaluation than their non-LGBTQ peers \cite{cech2021systemic,cech2020lgbt}. Because information about sexual orientation is not present in a scholar's name, efforts to increase opportunities for self-attestation are critical to future work seeking to evaluate impacts of sexual orientation on citation practices \cite{freeman2018lgbtq,langin2020lgbtq}. Similarly, trans and non-binary status cannot be predicted from names, and hence name-based assessments such as those we use here will not be able to reveal specific discrimination faced by these individuals due to their non-cis gender identity. Nevertheless, transgender and non-binary individuals exist in our dataset, and their name-(binary)gender associations are part of the probabilistic databases we employ. At the population scale, these individuals face the same name-based citation costs and payoffs as a cis gender scholar, which could be compounded by other costs and payoffs not accessible to our analysis. Of course, all scholars with marginalized intersecting identities are likely to face compounded citation costs, as indeed supported by a recent study that evaluated the intersection of gender and race/ethnicity \cite{Bertolero2020}. We look forward to further work that could complexify our understanding of citation practices, and their ethics, along these diverse lines of difference. 

\section{Conclusion}

Our work represents an analysis of the gendered demographics and citation practices of papers published in 35 physics journals over the last 25 years.
We found that citation practices skew toward the over-citation of man-authored papers (and the under-citation of woman-authored papers), and that citation behavior is heavily influenced by citing actor (the predicted gender makeup of the citing author team), citing venue (the subfield and journal in which the citation appears), and citing form (whether the citation pertains to work with which authors are likely to be familiar).
Our results highlight correlates with citation imbalance at the population level that are suggestive of actions that individuals and journals can take immediately to mitigate citation imbalance in physics, including (for individuals) acknowledging and disclosing the demographics of constructed reference lists, considering expanding said reference lists, and being mindful when citing outside one's specialty; and (for journals) increasing representation of woman authors, relaxing reference list length limits, and encouraging disclosure in the form of citation diversity statements.
Thoughtful engagement with citation imbalance at all levels of the publishing hierarchy will contribute to the inclusion of women in the current and future story (and reality) of scientific progress.
We invite readers to consider these and other ways they can make visible the contributions of all scholars to the scientific endeavor.

\section{Citation diversity statement}
Recent work has identified a bias in citation practices such that papers from women and other marginalized scholars in STEM are under-cited relative to the number of such papers in the field. 
Here we sought to proactively consider choosing references that reflect the diversity of the field in thought, form of contribution, gender, and other factors. 
We use databases that store the probability of a name being carried by people of different genders to mitigate our own citation bias at the intersection of name and identity. 
Based on the databases used, the set of names assigned the ``woman'' label will contain a predominance of women and the set of names assigned the ``man'' label will contain a predominance of men, but both sets may also contain other genders. 
By this measure (and excluding self-citations to the first and last authors of our current paper, and papers whose authors' first names could not be determined), our references contain 44.58\% woman(first)/woman(last), 19.28\% man/woman, 15.66\% woman/man, and 20.48\% man/man categorization. 
This method is limited in that names, pronouns, and social media profiles used to construct the databases may not, in every case, be indicative of gender identity. 
Furthermore, probabilistic studies of names cannot be used to detect citation costs that are specific to intersex, non-binary, or transgender people who are out to a large number of their colleagues. 
We look forward to future work that could help us to better understand how to support equitable practices in science.

\section{Acknowledgments}
E.G.T, L.C.B. and D.S.B. are supported by the National Science Foundation Materials Research Science and Engineering Center at University of Pennsylvania (NSF grant DMR-1120901). J.Z.K. is supported by a National Science Foundation Graduate Research Fellowship. C.W.L. is supported by a James S. McDonnell Foundation Postdoctoral Fellowship. S.C.S. acknowledges support from the University Scholars program at the University of Pennsylvania. P.S. is supported by the Swartz Foundation. P.Z. and D.S.B also acknowledge support from the Center for Curiosity. 

\section{References}
\bibliographystyle{unsrt}
\bibliography{DIVERSITY,mitigation,NeuroCite}
\end{document}